\documentclass[journal]{IEEEtran}
\usepackage{easyReview}
\usepackage{comment}

\usepackage{multirow}
\usepackage{graphicx}
\usepackage{float}
\usepackage{subcaption}
\usepackage{epstopdf}
\usepackage{amsmath}
\usepackage{amsfonts}
\usepackage{algorithm}
\usepackage{algorithmic}
\usepackage{makecell}
\usepackage{booktabs}
\setcellgapes{3pt}

\setcellgapes{3pt}

\usepackage[justification=centering]{caption}
\usepackage{color}
\usepackage{multirow}
\usepackage{tabularx}
\usepackage{amssymb}

\makeatletter

\newcommand{\Rmnum}[1]{\expandafter\@slowromancap\romannumeral #1@}
\makeatother

\usepackage{cite}

\ifCLASSINFOpdf
\else
\fi
\begin{document}
\bstctlcite{IEEEexample:BSTcontrol}
%
\title{M4SC: An MLLM-based Multi-modal, Multi-task and Multi-user Semantic Communication System}

\author{Feibo Jiang, \textit{Senior Member, IEEE}, Siwei Tu, Li Dong,  Kezhi Wang, \textit{Senior Member, IEEE}, Kun Yang, \textit{Fellow, IEEE}, Cunhua Pan,  \textit{Senior Member, IEEE}
}
\markboth{Submitted for Review}%
{Shell \MakeLowercase{\textit{et al.}}: Bare Demo of IEEEtran.cls for IEEE Journals}
%



\maketitle


\begin{abstract}
Multi-modal Large Language Models (MLLMs) are capable of precisely extracting high-level semantic information from multi-modal data, enabling multi-task understanding and generation. This capability facilitates more efficient and intelligent data transmission in semantic communications. In this paper, we design a tailored MLLM for semantic communication and propose an MLLM-based Multi-modal, Multi-task and Multi-user Semantic Communication (M4SC) system. First, we utilize the Kolmogorov-Arnold Network (KAN) to achieve multi-modal alignment in MLLMs, thereby enhancing the accuracy of semantics representation in the semantic space across different modalities. Next, we introduce a multi-task fine-tuning approach based on task instruction following, which leverages a unified task instruction template to describe various semantic communication tasks, improving the MLLM's ability to follow instructions across multiple tasks. Additionally, by designing a semantic sharing mechanism, we transmit the public and private semantic information of multiple users separately, thus increasing the efficiency of semantic communication. Finally, we employ a joint KAN-LLM-channel coding strategy to comprehensively enhance the performance of the semantic communication system in complex communication environments. Experimental results validate the effectiveness and robustness of the proposed M4SC in multi-modal, multi-task, and multi-user scenarios.

\end{abstract}


\IEEEpeerreviewmaketitle

\section{Introduction}
\begin{figure}[htpb]
	\centering
	\includegraphics[width=8.5cm]{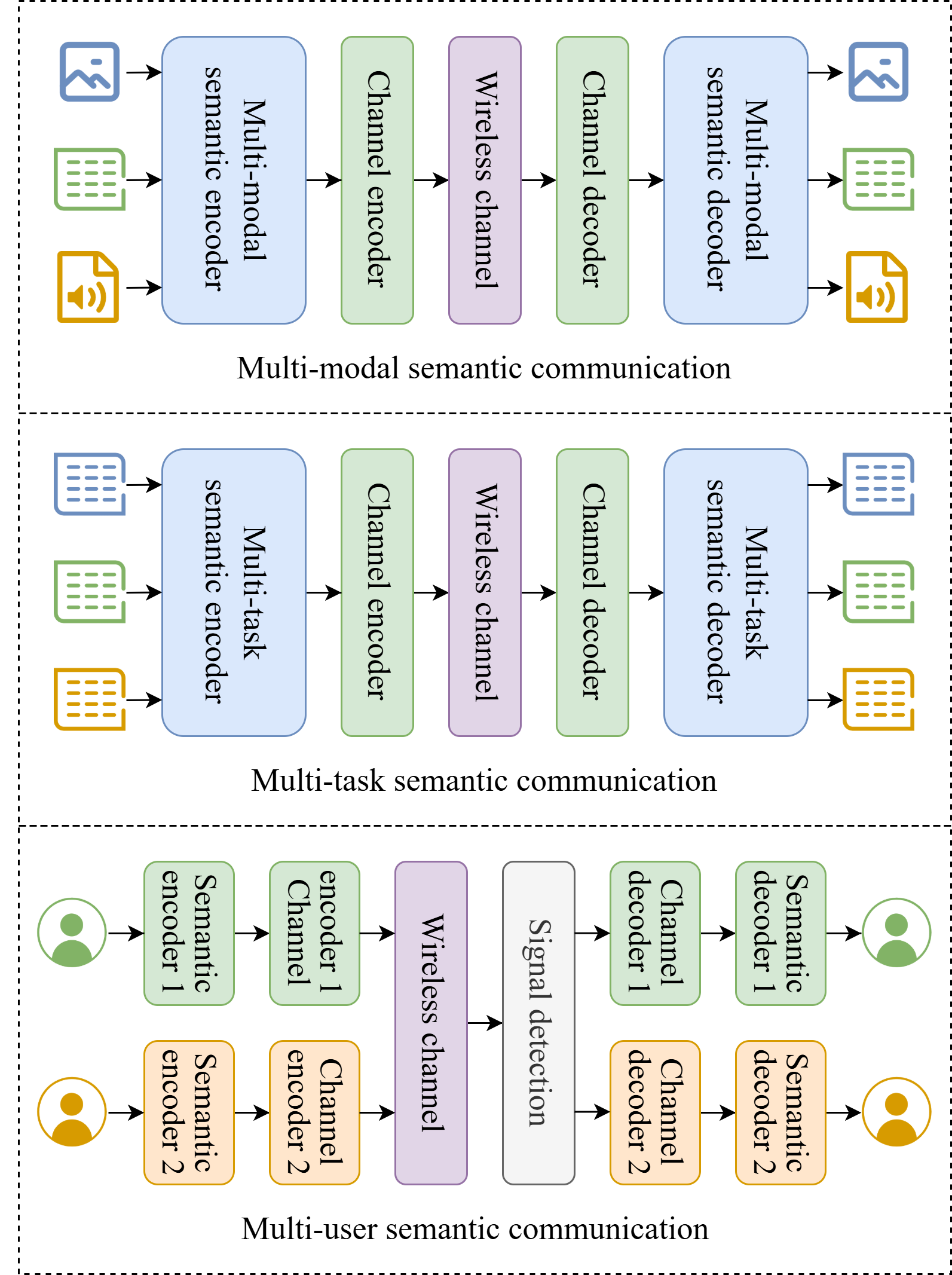}
	\caption{The structure of three different semantic communications.}
	\label{fig:threeKind}
\end{figure}

Semantic communication is a communication paradigm centered around conveying the ``meaning" of source data, with its primary goal being to ensure that the receiver understands the sender's intent, rather than simply receiving precise bit-level data. A semantic communication system typically consists of several key components: (1) \emph{Semantic encoder}, which processes the source data at the semantic level, extracting and encoding the core semantic information of the source data rather than transmitting the raw data directly. (2) \emph{Channel encoder}, which further processes the semantic-encoded information to adapt them to the transmission characteristics of the communication channel, such as noise and attenuation. (3) \emph{Channel decoder}, which decodes the signal received through the physical channel and restores the semantic information. (4) \emph{Semantic decoder}, which decodes the semantic information and ultimately reconstructs the data into a form that the user can understand. (5) \emph{Knowledge base}, which contains various background and prior information that aids the system in better understanding and recovering the true meaning of the source data during the encoding and decoding process. The core objective of the entire semantic communication is to achieve efficient and accurate semantic transmission, surpassing the traditional reliance on bitstreams in communication systems, and therefore performing more effectively in resource-constrained environments \cite{yang2022semantic}. In recent years, with the advancement of deep learning, semantic communication has gradually become a key research focus in 6G. It is evolving towards multi-modal, multi-task, and multi-user development trends, as shown in Fig. \ref{fig:threeKind}.

\subsection{Multi-modal semantic communication}
\subsubsection{Definition}
Multi-modal semantic communication refers to communication systems capable of simultaneously processing and transmitting multiple types of modal data, such as text, images, and video. The core objective is to extract semantic information from different modalities to enable the efficient sharing and expression of cross-modal information. Multi-modal semantic communication leverages the semantic relationships between modalities in both time and space to ensure high-quality semantic understanding, creating more comprehensive semantic expressions and thereby enhancing the integrity of the transmitted information \cite{jiang2024large}.

\subsubsection{Challenges}
Different modalities inherently differ in their data forms and semantic representations, and their feature spaces and semantic distributions are not aligned. For example, text is based on vocabulary and syntactic structures, images rely on pixels and visual features, and audio and video consist of time-series signals. This heterogeneity makes it exceptionally challenging to achieve a unified understanding across modalities at the semantic level. Cross-modal alignment is a key technology in multi-modal semantic communication, where data from different modalities are mapped into a shared semantic space through semantic analysis or feature extraction. This enables the modalities to understand one another and ultimately achieves consistent semantic representation.

\subsection{Multi-task semantic communication}
\subsubsection{Definition}
Multi-task semantic communication refers to a communication architecture that supports the simultaneous transmission of semantic information for multiple tasks. Its goal is to optimize the semantic requirements of different tasks, enabling knowledge sharing and collaborative communication across tasks through a unified semantic representation. Multi-task semantic communication allows for the dynamic adjustment of communication strategies to accommodate varying task priorities and service quality demands\cite{zhang2024unified}.

\subsubsection{Challenges}
The semantic spaces and objectives of different tasks may differ or even conflict, requiring the model parameters in multi-task semantic communication to be balanced across tasks. For instance, when the system learns a new task, adjustments to the model parameters may result in the loss of important semantic information from previous tasks, leading to the phenomenon of ``catastrophic forgetting" in multi-task learning. Therefore, during the training of multi-task semantic communication, the system typically needs to optimize multiple tasks in the semantic and channel encoders, ensuring that the learning of new tasks and adaptation to new environments can proceed without disrupting the performance of existing tasks.

\subsection{Multi-user semantic communication}
\subsubsection{Definition}
Multi-user semantic communication refers to communication systems that support the simultaneous transmission of semantic information for multiple users. It is primarily based on traditional multiple access technologies that divide physical resources (such as frequency division multiple access, time division multiple access, and code division multiple access). The objective is to optimize the semantic transmission efficiency among users through joint modeling and transmission of user semantic information, thereby significantly reducing bandwidth consumption while ensuring fairness and service quality \cite{xie2021task}.

\subsubsection{Challenges}
The key challenges in multi-user semantic communication research lie in how to effectively allocate resources and optimize strategies to meet the diverse needs of multiple users while maximizing the overall semantic transmission efficiency of the system. Therefore, multi-user semantic communications must scientifically allocate limited communication resources to avoid resource wastage and interference among users. Additionally, they must fully consider the collaborative mechanisms among users, such as reducing information redundancy through shared semantic information or collaborative encoding/decoding, in order to achieve global optimization of semantic transmission.

\subsection{Main contributions}
Multi-modal Large Language Models (MLLMs) represent the cutting-edge research direction in generative artificial intelligence and demonstrate significant potential in semantic information processing. By leveraging cross-modal semantic understanding, MLLMs integrate heterogeneous data such as text, images, and video, enabling multimodal information comprehension at the semantic level. Through the establishment of shared semantic representations across multiple tasks, MLLMs optimize task performance in the semantic encoder, enabling robust multi-task learning. Additionally, MLLMs capture the similarities and differences in semantics between different users in the semantic space, utilizing collaborative mechanisms to improve the overall transmission efficiency of semantic communication. Therefore, in this paper, we design a tailored MLLM for semantic communication and propose an MLLM-based Multi-modal, Multi-task, and Multi-user Semantic Communication (M4SC) system. Our contributions can be summarized as follows:

\subsubsection{KAN-based multi-modal alignment}
To address the challenge of semantic alignment in multi-modal semantic communication, we propose a cross-modal projection alignment method based on the Kolmogorov-Arnold Network (KAN). This method decomposes visual semantic information into several low-dimensional nonlinear subspaces through a set of learnable activation functions. It then dynamically learns the optimal combination of visual features, aligning the visual semantic information more smoothly and accurately with textual semantic information of Large Language Models (LLMs) in the semantic space. This significantly improves the accuracy and efficiency of multi-modal semantic alignment.

\subsubsection{Multi-task instruction fine-tuning}
To overcome the challenge of catastrophic forgetting in multi-task semantic communication, we propose a multi-task fine-tuning method based on the task instruction following. By using natural language instructions to uniformly describe the objectives, inputs, and outputs of different tasks, we fine-tune the MLLM to enhance its multi-task learning capability and generalization ability. The design of the instructions emphasizes task clarity, semantic consistency, and scalability, enabling the model to understand the latent relationships between tasks. This approach significantly improves the robustness of semantic communication in multi-task scenarios, allowing it to dynamically adapt to task demands while reducing interference between tasks.

\subsubsection{Multi-user semantic sharing transmission}
To achieve efficient collaboration among multi-user communication, we establish a shared semantic space between users, where the semantic information of different users is compared and classified. Semantic information that is identical or similar is merged to form public information. On the sender side, we transmit the public and private information separately, making more efficient use of spectrum resources. On the receiver side, users reconstruct the original data by combining the public information with their respective private information, thus enhancing the overall semantic transmission efficiency of the system.

The remainder of this paper is organized as follows: Section \Rmnum{2} introduces the tailored MLLM for semantic communication systems; Section \Rmnum{3} presents a detailed introduction to the proposed M4SC; Section \Rmnum{4} details the experimental setup and results; 
and Section \Rmnum{5} provides a summary and discussion of the paper.

\section{MLLM for Semantic Communication Systems}
In recent years, breakthroughs in LLMs for natural language understanding and text generation have injected new vitality into semantic communication. By harnessing the powerful semantic reasoning and generative capabilities of LLMs, semantic communication can achieve higher-level information transmission. MLLMs are an extension of LLMs, developed to handle and understand data from multiple modalities, including text, images, and video. MLLMs inherit several prominent features of LLMs, such as zero-shot learning, In-Context Learning (ICL), Chain-of-Thought (CoT), and instruction following \cite{jiang2024large2}. Notable examples of MLLMs include GPT-4o, Claude 3, and LLaVA. 
By introducing cross-modal semantic analysis capabilities, MLLMs can map semantic information from different modalities into a unified semantic space, further enhancing the depth and breadth of semantic expression, and becoming a key driving force in the future development of semantic communications.

\subsection{Advantages of MLLM in semantic communication}

\subsubsection{High-precision multi-modal alignment}
MLLMs, through large-scale pre-training, map the semantic information of data from different modalities into a shared semantic space. In this semantic space, common features between modalities can be captured and associated. For example, the description of a ``red apple" in text can be naturally aligned with the corresponding visual semantic information in an image. This unified semantic space provides a solid foundation for the coordination and fusion of multi-modal data, making it one of the core technologies for multi-modal alignment. To achieve high-precision alignment, MLLMs typically employ a cross-modal projector to transform visual semantic information into textual semantic representations that are compatible with LLMs. By training the cross-modal projector, MLLMs can learn the mapping rules between text and visual semantic information. This deep cross-modal understanding enables MLLMs to efficiently perform semantic encoding and decoding between multi-modal data in semantic communications.

\subsubsection{Superior robustness in multi-task learning}
The core challenge in multi-task semantic communication is enabling the system to quickly adapt to the requirements of different tasks. MLLMs, with their powerful generalization capabilities and flexibility, provide effective support for overcoming this challenge. Through large-scale pre-training and instruction fine-tuning, MLLMs exhibit exceptional zero-shot and few-shot learning abilities. Consequently, MLLMs inherently possess a fundamental understanding of various tasks and can rapidly adapt to new task requirements under zero-shot or few-shot conditions using advanced learning methods such as ICL, CoT, and Retrieval-Augmented Generation (RAG). This ability significantly reduces the training and updating costs of semantic communication systems.

\subsubsection{High-efficiency multi-user semantic transmission}
In multi-user semantic communication, precise semantic expression and comparison among users are key to achieving efficient communication. To achieve this, the system needs to compare and extract similar semantic information from a shared semantic space across multiple users, minimizing information redundancy while ensuring the integrity and consistency of user semantics. In MLLM-based semantic communications, the system can leverage their powerful semantic understanding and encoding capabilities to accurately identify both the commonalities and differences in the semantic information of different users, thereby improving overall communication efficiency.

\subsection{The tailored MLLM}
\begin{figure}[htpb]
	\centering
	\includegraphics[width=8.5cm]{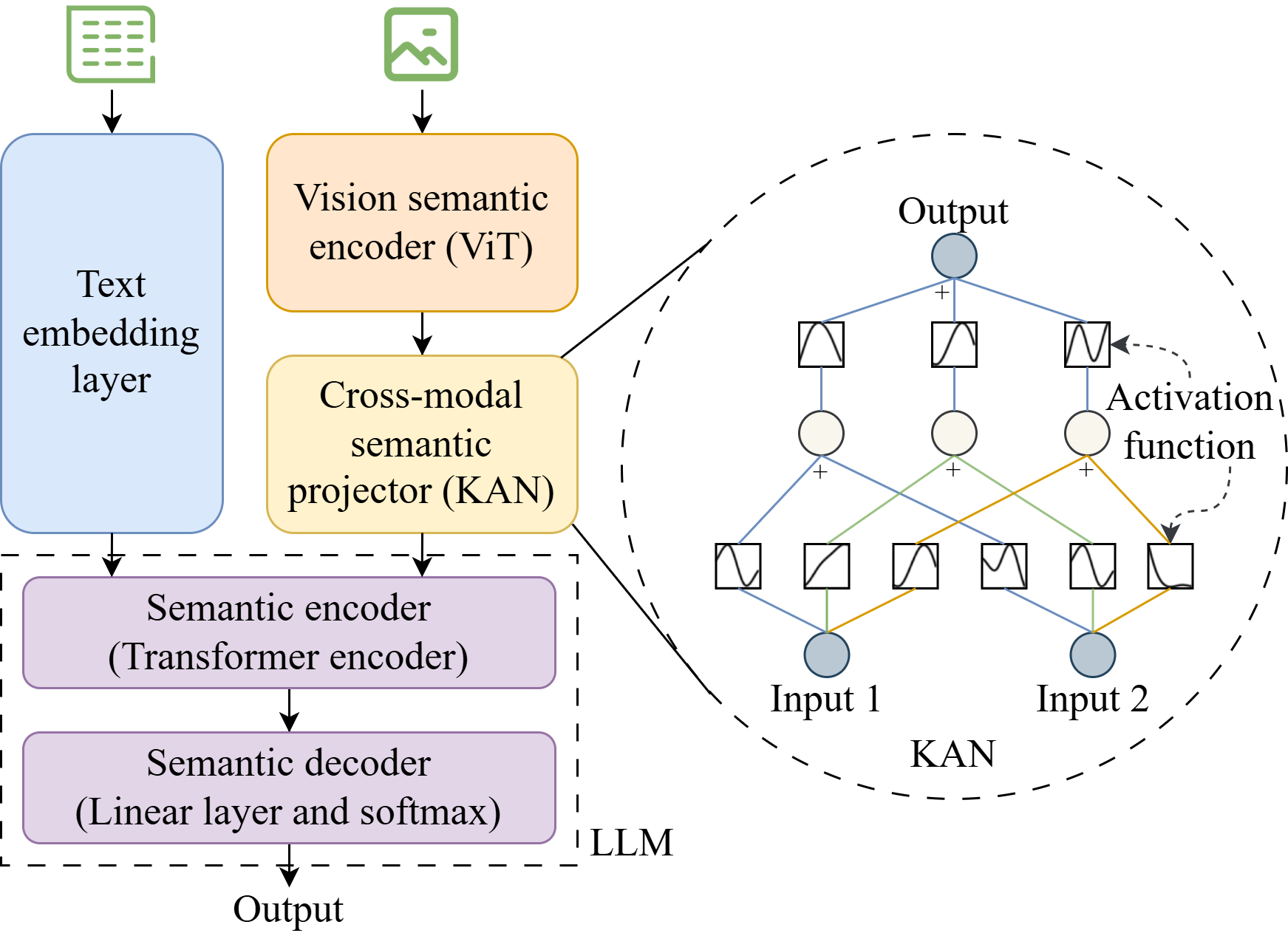}
	\caption{The structure of the proposed MLLM.}
	\label{fig:arch}
\end{figure}

Due to the advantages of MLLMs, we design a tailored MLLM for semantic communications. First, it is a lightweight MLLM architecture that supports semantic communication for both image and text modalities on resource-constrained devices. Second, it is an extension of existing LLMs, requiring minimal computational resources for multi-modal training, which facilitates subsequent joint encoding for semantic communications. The structure of the proposed MLLM is shown in Fig. \ref{fig:arch}.

\subsubsection{Vision semantic encoder}
The vision semantic encoder in the MLLM is primarily responsible for converting image data into visual semantic information. It employs visual processing architectures such as Convolutional Neural Networks (CNNs) or Vision Transformers (ViT) to extract visual features from images through convolutional kernels or self-attention mechanisms, and then convert these features into precise visual semantic information. In the proposed MLLM, we use the pre-trained Siglip ViT as the vision semantic encoder \cite{zhai2023sigmoid}. Siglip, trained through self-supervised contrastive learning on both text and image data, enables the ViT structure to efficiently and accurately extract semantic information from images.

\subsubsection{Cross-modal semantic projector}
The cross-modal semantic projector in the MLLM is responsible for aligning visual semantic information with the textual semantic space of the LLM. It transforms the visual semantic information extracted by the vision semantic encoder into a vector compatible with the textual semantic information, using a nonlinear transformation. This process allows the image and text to be compared and fused in the same semantic space, thereby achieving cross-modal information integration and alignment. In the proposed MLLM, we use the KAN as the cross-modal semantic projector. KAN utilizes a set of learnable spline functions as activation functions, enhancing the model's ability to capture subtle patterns in the semantic space, enabling smoother and more precise alignment of visual semantic information with the textual semantic information \cite{liu2024kan}.

\subsubsection{LLM-based semantic encoder}
The text embedding layer of the LLM converts text data into textual semantic information, which is then fused with the visual semantic information from the cross-modal semantic projector and input into the LLM's encoder. This allows the LLM to not only understand the textual content but also to combine visual semantic information for more accurate reasoning and generation. The LLM-based semantic encoder adopts a Transformer encoder architecture, utilizing stacked self-attention layers and fully connected layers to perform high-quality multimodal semantic encoding.

\subsubsection{LLM-based semantic decoder}
The LLM-based semantic decoder employs a linear layer followed by a softmax layer to convert the multi-modal semantic information encoded by the semantic encoder into a probability distribution for the generated content. It then selects the most probable content for semantic decoding, producing the final output. In the proposed MLLM, we use Google's Gemma2-2b-it \cite{team2024gemma} as the LLM-based semantic encoder and decoder. With only 2 billion parameters, Gemma2-2b-it offers a lightweight design and advantages for multitask generation. It operates efficiently in resource-constrained semantic communication systems and performs excellently in various text generation tasks such as question answering, text summarization, and logical reasoning.

\section{MLLM-based Multi-modal, Multi-task and Multi-user Semantic Communications}
\begin{figure*}[htpb]
	\centering
	\includegraphics[width=17cm]{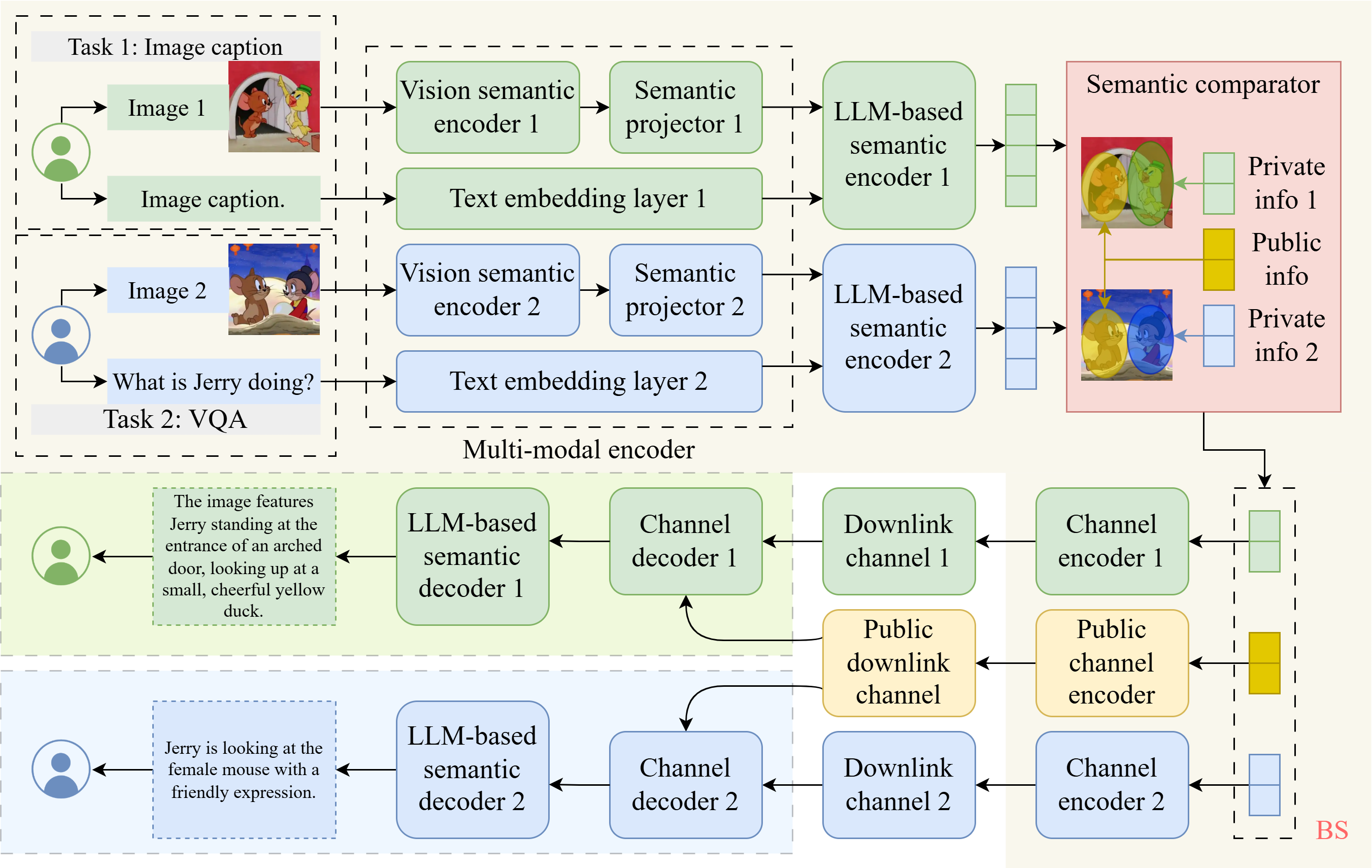}
	\caption{The structure of the proposed M4SC.}
	\label{fig:ours}
\end{figure*}
Based on the tailored MLLM, we design a multi-modal, multi-task, and multi-user semantic communication system, as illustrated in Fig. \ref{fig:ours}. In M4SC, we consider a multi-user semantic communication scenario from the Base Station (BS) to the users. As a specific example, at the BS, two users perform image captioning (Task 1) and Visual Question Answering (VQA) (Task 2), both of which are multimodal tasks, providing their respective image and text inputs. The multimodal encoder, consisting of a text embedding layer, a vision semantic encoder, and a cross-modal semantic projector, simultaneously aligns the inputs from the image and text modalities, which are then semantically encoded by the LLM-based semantic encoder. Subsequently, the semantic comparator divides the semantic information of the two users into public semantic information (e.g., the semantic information represented by "Jerry" in the two images) and private semantic information (e.g., the semantic information represented by "yellow duck" in Image 1 and "female mouse" in Image 2). The public and private semantic information is then encoded and transmitted by the public channel encoder and the user-specific channel encoder, respectively. On the receiver side, users reorganize the received public and private semantic information and sequentially use the channel decoder and the LLM-based semantic decoder to obtain the output, thus completing the semantic communication tasks for each user.

\subsection{KAN-based cross-modal alignment}
To achieve more precise multi-modal understanding and cross-modal alignment, we use the KAN as the cross-modal semantic projector in the M4SC. Based on the Kolmogorov-Arnold representation theorem, KAN offers an efficient and accurate neural network architecture for representing nonlinear functions. It is capable of more precisely capturing the multi-dimensional semantic information of input data while maintaining computational efficiency and ensuring the accuracy of the output results. The workflow of KAN is as follows:

\subsubsection{Activation and decomposition Stage}
In this stage, the input signals are passed through a set of learnable activation functions, each of which is independently adjustable. These activation functions, constructed using spline functions, transform the complex multi-modal semantic information into simple polynomial combinations. The activation values for each input node are computed multiple times and sequentially aggregated in the order of the activation functions.

\subsubsection{Aggregation and Mapping Stage}
After the activation values are aggregated, the sum of these values serves as the output for the current layer. The output node count corresponds to the number of times each input node is activated. This sequential aggregation process enables the network to establish more intricate mapping relationships between layers, allowing for more accurate representation and alignment of multi-modal data, thus capturing deeper associations between different modalities. The network’s hierarchical structure is further enriched and optimized through iterative activations, improving overall performance in complex data mapping tasks.

\subsection{Multi-task instruction fine-tuning}
To enhance the performance and generalization ability of the multi-task semantic communication, we propose a multi-task fine-tuning method based on the task instruction following. This approach unifies the description of the objectives and input-output formats for different tasks through natural language instructions and employs a multi-task mixed training strategy to optimize the model’s generalization capability. The model can thus understand the potential relationships between tasks, leading to an overall improvement in multi-task performance \cite{wei2021finetuned}. 

A complete task instruction consists of the following components: Instruction, Input, Output, and Metadata. The Instruction clearly describes the content and objective of the task; the Input provides the contextual information for the instruction; the Output provides a high-quality result based on the instruction and input that meets the task requirements; and the Metadata (optional) includes additional details such as task notes. These elements work together to enable the M4SC to accurately understand and perform semantic communication tasks.

For example, in Fig. \ref{fig:ours}, the task instruction of the image caption is as follows: \emph{\textbf{Instruction}: ``Please carefully observe the provided image and identify the main elements, including the scene (e.g., indoor, outdoor, city, nature, etc.), key objects (e.g., animals, people, buildings, tools, etc.), and any actions or activities (e.g., running, eating, reading, etc.). Based on this information, generate a natural language description that covers the key details of the image. Ensure the description is fluent, grammatically correct, and avoids excessive speculation, only describing what can be clearly observed in the image." \textbf{Input Image}: \textless Image 1\textgreater \ \textbf{Input Text}: ``Image caption." \textbf{Output}: ``The image features Jerry standing at the entrance of an arched door, looking up at a small, cheerful yellow duck."}

The task instruction of VQA is as follows: \emph{\textbf{Instruction}: ``Please observe the provided image and read the question carefully to understand its specific requirements. Based on the content of the image, answer the question, ensuring that the answer is based on clearly observable information. If the question involves specific details (e.g., quantity, color, location), provide precise answers. If the question cannot be answered based on the image, respond with: `Cannot obtain an answer from the image.' Finally, provide a concise and clear answer that fully satisfies the semantic requirement of the question." \textbf{Input Image}: \textless Image 2\textgreater \   \textbf{Input Text}: ``What is Jerry doing?" \textbf{Output}: ``Jerry is looking at the female mouse with a friendly expression."} 

\subsection{Semantic shared transmission for multi-user}
Due to the potential high similarity in the metadata of data across different users at the semantic level, this phenomenon provides a theoretical foundation for the comparison and sharing of semantic information between users. In M4SC, we consider a multi-user semantic communication scenario from the BS to users, where similar or identical semantic information vectors from different users are merged to form a public semantic signal for transmission. This approach makes more efficient use of spectrum resources, improving the overall semantic transmission efficiency. The process of semantic shared transmission is as follows:


\subsubsection{Shared semantic space mapping}
At the BS, we convert each user's input source data into multimodal semantic information and map them into a unified semantic space using the cross-modal semantic projector.

\subsubsection{Semantic comparison}
In the semantic space, we perform a detailed, token-level comparison of the semantic information from different users, calculating their similarity. The statistical characteristics of the semantic information vectors (e.g., mean, variance) are compared to a threshold to determine whether there are identical or similar semantics \cite{zhang2023model}.

\subsubsection{Public and private semantic partitioning}
Identical or similar semantics between users are merged to form public semantic information, while the unique, personalized semantic information (private semantic information) of each user is retained.

\subsubsection{Semantic shared transmission}
During wireless transmission, the encoded public semantic information is sent through a dedicated public channel and broadcast to all users. The encoded private semantic information is transmitted to the corresponding receiving users through their assigned channels.

\subsubsection{Semantic reconstruction}
Each receiving user combines the private semantic information with the public semantic information they have received, then performs channel decoding followed by semantic decoding to complete the semantic communication task.

\subsection{Training process of the semantic communication system}

\begin{figure}[htpb]
	\centering
	\includegraphics[width=8.5cm]{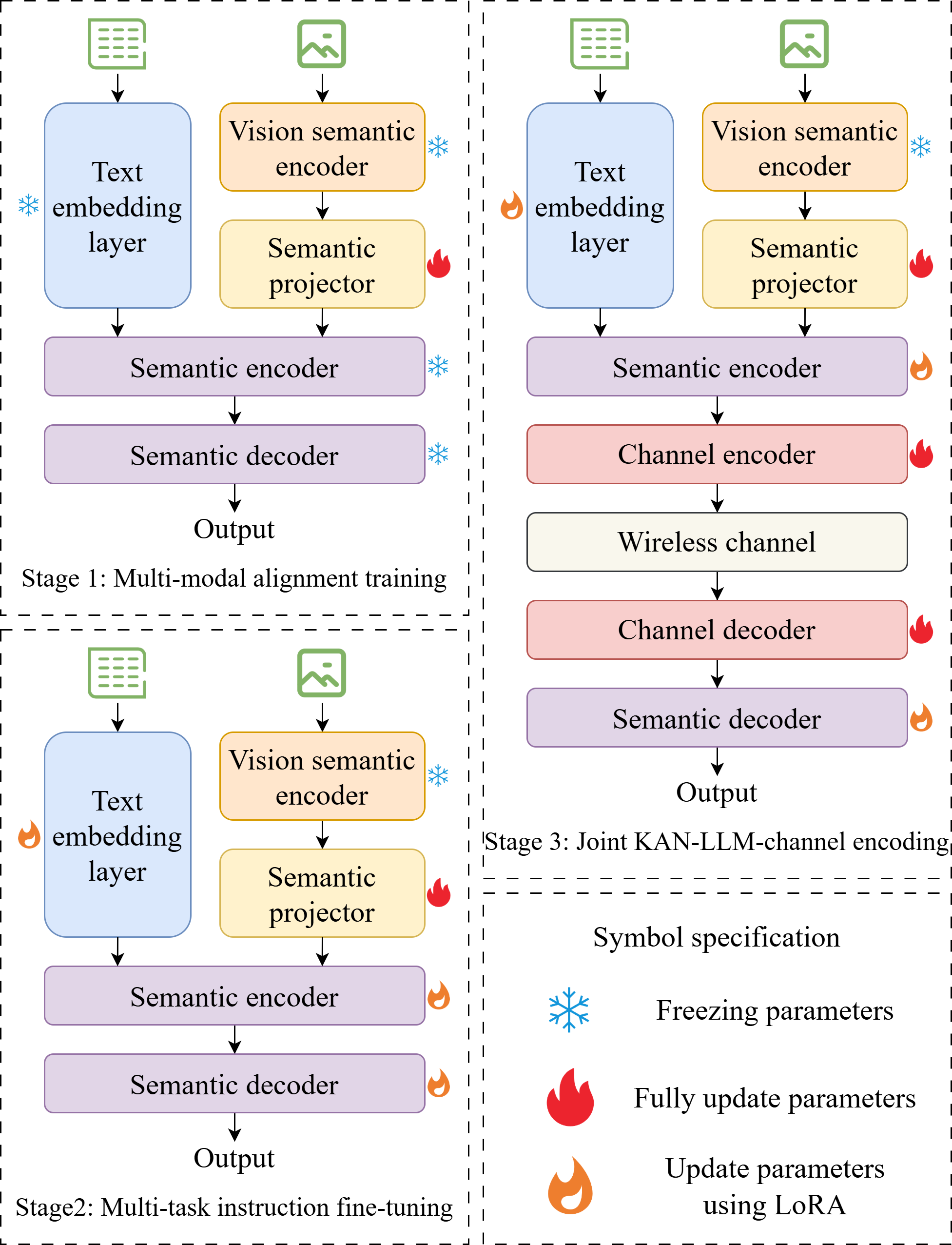}
	\caption{The three-stage training process of M4SC.}
	\label{fig:train}
\end{figure}

To further optimize the performance of the M4SC, we customize a three-phase training process that includes two phases of MLLM training and a third phase for joint encoding of the semantic communication system, as shown in Fig. \ref{fig:train}.

\subsubsection{Multi-modal alignment training}
The first phase focuses on training the KAN-based cross-modal semantic projector. Specifically, the goal of this phase is to train an efficient tokenizer that adapts to the visual modality for a frozen LLM. This tokenizer is designed to map image semantic information into textual semantic information compatible with the LLM, thereby enhancing the model's understanding of image modality inputs. This training phase ensures that the cross-modal semantic projector can efficiently adapt to the image data required in multi-modal semantic communication without needing to train the LLM.

\subsubsection{Multi-task instruction fine-tuning}
In the second phase, we no longer freeze the LLM but continue training both KAN and LLM parameters. The purpose of this phase is to strengthen the system's multi-modal understanding capabilities while improving the model’s ability to follow instructions for different tasks. This joint optimization approach not only ensures the accuracy of MLLM in understanding multi-modal tasks but also enhances its robustness when executing multiple tasks simultaneously. Additionally, to accelerate computation and reduce resource consumption, we apply Low-Rank Adaptation (LoRA) for efficient fine-tuning of the LLM.

\subsubsection{Joint KAN-LLM-channel encoding}
In the third phase, we further integrate the optimized KAN and LLM with the channel encoder-decoder for joint training. The focus of this phase is to improve the overall performance of the semantic communication system. In the joint KAN-LLM-channel encoding process, we allow the parameters of KAN and the channel encoder-decoder to undergo full updates to accommodate the complex channel conditions. We also use LoRA to fine-tune the LLM. This strategy significantly reduces training costs while maintaining model performance, making it more suitable for joint encoding training in M4SC.

\section{Simulation Results}

\subsection{Experimental setup}
The specific experimental setup is as follows: The vision semantic encoder of the MLLM is a pre-trained Siglip ViT, with 0.3 billion parameters. The cross-modal semantic projector is a two-layer KAN, with 76 million parameters. The LLM used in the MLLM is Gemma2-2b-it, with 2 billion parameters. The channel encoder consists of a linear layer, with an input feature dimension of 2304 and an output feature dimension of 512. To maintain information consistency, the channel decoder adopts a structure that is the inverse of the channel encoder. The wireless channel model adopts similar settings to those presented in \cite{jiang2024large}.


In multi-modal alignment training, 1.2 million image-text description pairs are used, with datasets including LLaVA Images and ALLaVA Caption \cite{chen2024allava}. Multi-task instruction fine-tuning uses 1.5 million single-turn or multi-turn image-text dialogue pairs, with datasets including COCO train2017, GQA, OCR-VQA, TextVQA, VisualGenome part1/part2, ShareGPT4V-100K, LAION GPT4V, ALLaVA Instruction, DocVQA, ChartQA, DVQA, and AI2D \cite{li2024mini}. The joint KAN-LLM-channel encoding uses the CLEVR dataset \cite{johnson2017clevr}. Moreover, the AdamW optimizer and cosine learning rate scheduling are used in the training process.


Training and testing are conducted on a server equipped with an Intel Xeon CPU (2.6 GHz, 1007 GB RAM) and 8*NVIDIA A100 GPUs (80 GB SGRAM), with the operating environment being Python 3.10. The training framework used is PyTorch 2.0.1, and the training strategy employs DeepSpeed ZeRO-2.

\subsection{Evaluation results}
\subsubsection{Multi-modal and multi-task performance evaluation}
To assess the multi-modal performance of the M4SC, we compare it with DeepSC-VQA \cite{xie2022task} on the VQA task. The dataset used for the VQA task is CLEVR\cite{johnson2017clevr}. To evaluate the multi-task performance of the M4SC, we compared it with both DeepSC-VQA and DeepSC \cite{xie2021deep} across various downstream tasks, including VQA and text classification. The dataset used for text classification is the IMDB dataset. The evaluation metric for both experiments is accuracy.

\begin{figure}[htpb]
	\centering
	\includegraphics[width=8.5cm]{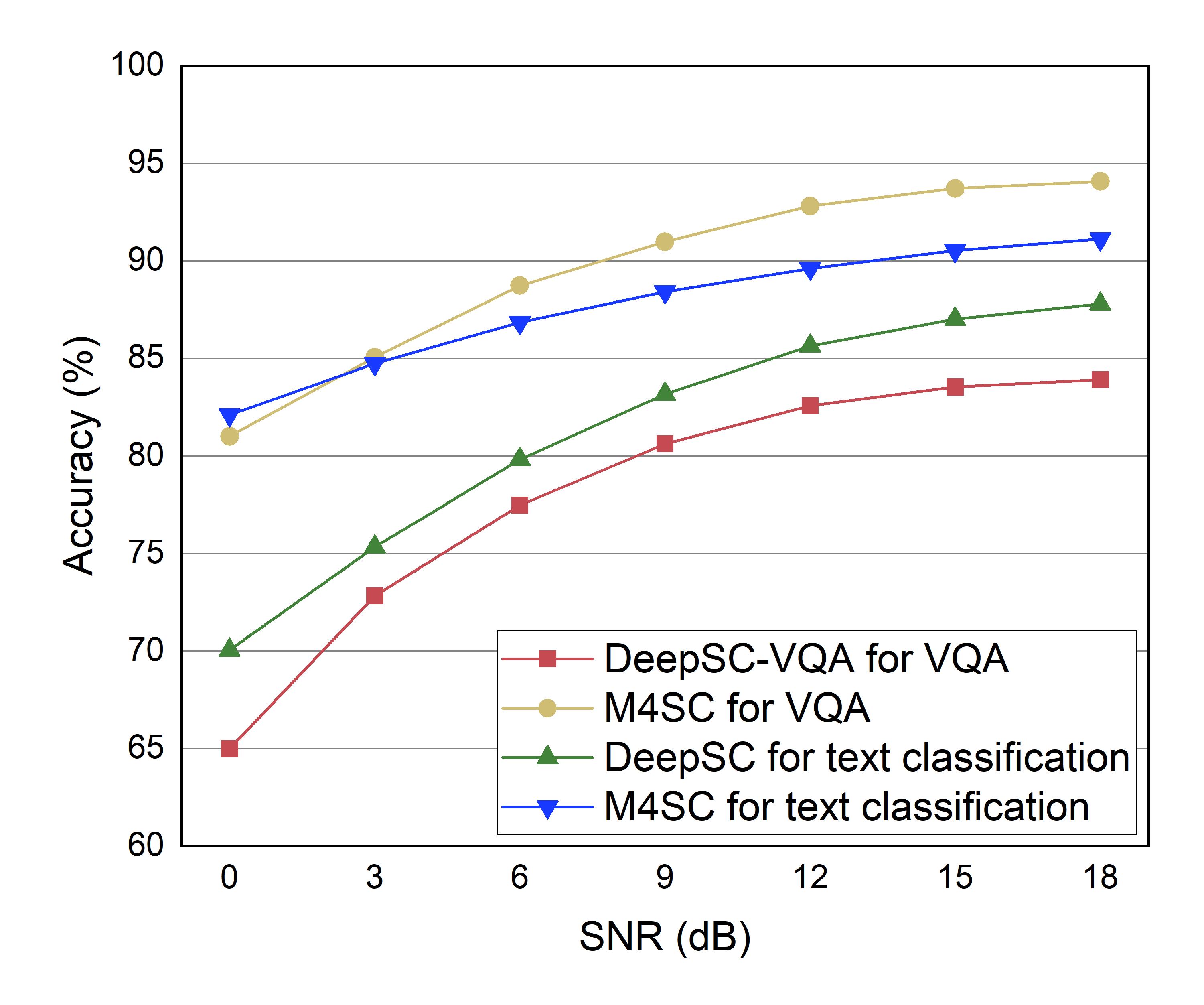}
	\caption{Comparison of multi-modal and multi-task performance.}
	\label{fig:p1}
\end{figure}

The performance comparison results for multi-modal and multi-task tasks are shown in Fig. \ref{fig:p1}. It can be observed that, whether for the multi-modal task of VQA or the unimodal task of text classification, M4SC consistently outperforms the semantic communication systems designed for single tasks. This demonstrates that the powerful multi-modal alignment and multi-task learning capabilities of MLLM enable it to excel in multi-modal and multi-task scenarios. Specifically, MLLM effectively maps data from different modalities into a unified semantic space and flexibly adapts to various semantic task requirements through multi-task instruction fine-tuning.

\subsubsection{Multi-user performance evaluation}
\begin{figure}[htpb]
	\centering
	\includegraphics[width=8.5cm]{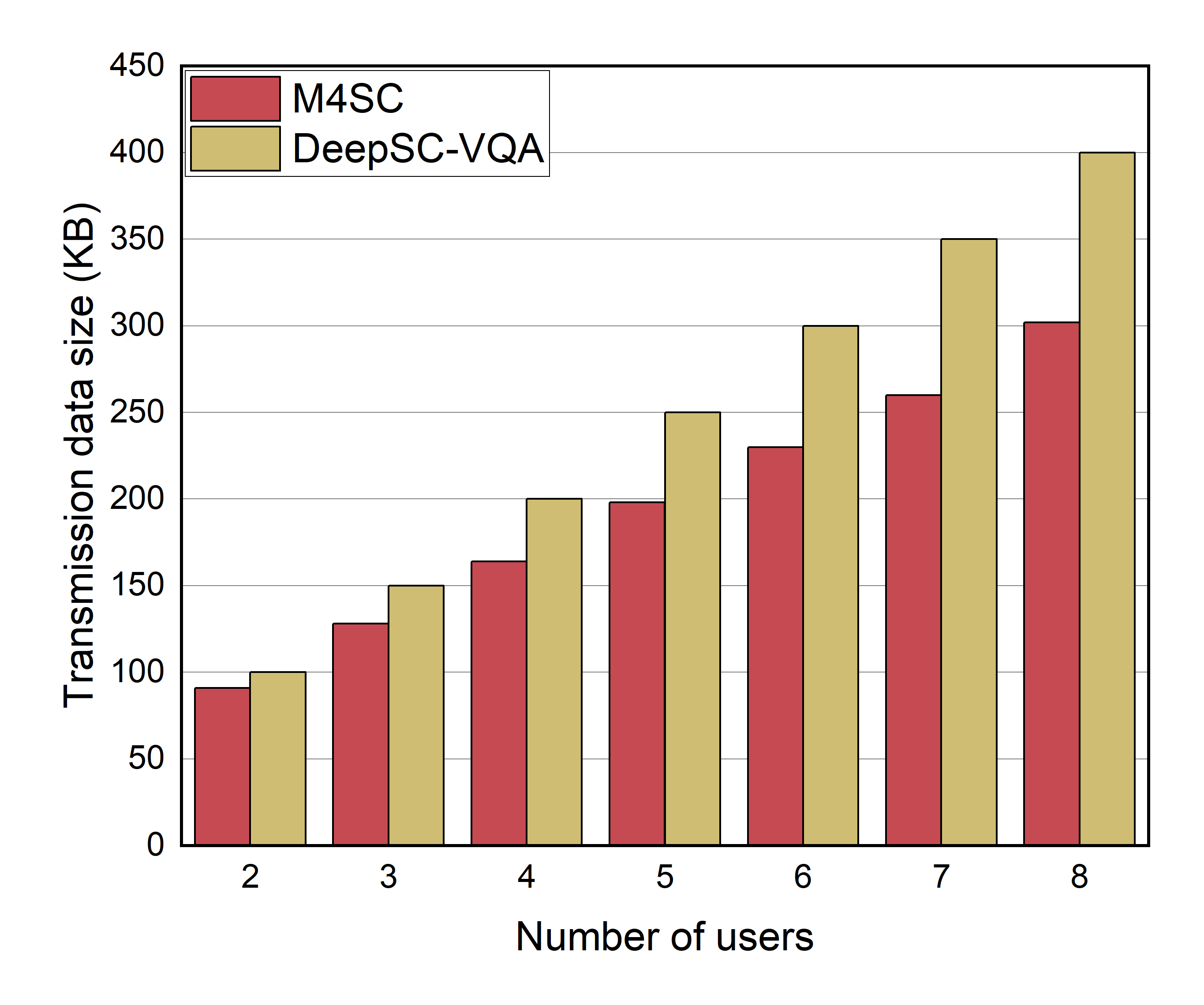}
	\caption{Comparison of multi-user performance.}
	\label{fig:p2}
\end{figure}

To evaluate the advantages of M4SC in multi-user communication, we compare the size of the transmitted data for the VQA task between M4SC and DeepSC-VQA under different numbers of users using CLEVR dataset \cite{johnson2017clevr}, as shown in Fig. \ref{fig:p2}. 

As shown in Fig. \ref{fig:p2}, with the increase in the number of users, M4SC demonstrates significant advantages in terms of the size of the transmitted data. Compared to the DeepSC-VQA, the proposed M4SC effectively reduces redundant data transmission through its semantic sharing mechanism. Specifically, when the number of users is small, there is little difference in data transmission between the two systems. However, as the number of users increases, the M4SC significantly reduces the overall transmission data size by extracting and transmitting public semantic information. This result indicates that the proposed semantic sharing transmission mechanism has a remarkable bandwidth optimization effect in multi-user scenarios, effectively improving the overall efficiency of the system.

\section{Conclusion}
This study presented a novel MLLM designed for semantic communication and introduced an MLLM-based multi-modal, multi-task and multi-user semantic communication system, named M4SC. The M4SC achieved precise alignment of multi-modal data in a shared semantic space by incorporating KAN as a cross-modal semantic projector, significantly enhancing the accuracy of multi-modal semantic representation. 
Furthermore, a multi-task instruction fine-tuning method was proposed, where unified instructions provide natural language descriptions for different tasks, thus improving the M4SC's multi-task learning. In multi-user scenarios, the M4SC employed a semantic sharing mechanism to separate the transmission of public and private semantic information, further enhancing the overall semantic transmission efficiency. Simulation experiments demonstrated that M4SC not only excels in multi-modal and multi-task semantic communication but also exhibits remarkable bandwidth optimization in multi-user semantic communication.


%

%




\bibliographystyle{ieeetran}
\bibliography{bare_jrnl_bobo}
\section*{Biographies}
\textbf{Feibo Jiang} (jiangfb@hunnu.edu.cn) received Ph.D. degree from the Central South University, China. He is currently an Associate Professor at Hunan Normal University, China.

\textbf{Siwei Tu} (tusiwei@hunnu.edu.cn) is currently pursuing the master’s degree with Hunan Normal University, China. 

\textbf{Li Dong} (Dlj2017@hunnu.edu.cn) received Ph.D. degree from the Central South University, China. She is currently an Associate Professor at Hunan University of Technology and Business, China.

\textbf{Kezhi Wang} (Kezhi.Wang@brunel.ac.uk) received Ph.D. degree from University of Warwick, U.K. in 2015. Currently he is a Senior Lecturer with the Department of Computer Science, Brunel University London, U.K.

\textbf{Kun Yang} (kunyang@essex.ac.uk) received his PhD from the Department of Electronic \& Electrical Engineering of University College London (UCL), U.K. He is currently a Chair Professor in the School of Computer Science \& Electronic Engineering, University of Essex, U.K.

\textbf{Cunhua Pan} (cpan@seu.edu.cn) received Ph.D. degrees from Southeast University, China, in 2015. 
He is a full professor in Southeast University, China.

\end{document}